# Distance Distributions for Real Cellular Networks

Siyi Wang, Weisi Guo[†], Mark D. McDonnell

Institute for Telecommunications Research, University of South Australia, Australia

[†]School of Engineering, The University of Warwick, United Kingdom

siyi.wang@mymail.unisa.edu.au, [†]weisi.guo@warwick.ac.uk, mark.mcdonnell@unisa.edu.au

*Abstract*—This paper presents the general distribution for the distance between a mobile user and any base station (BS). We show that a random variable proportional to the distance squared is *Gamma* distributed. In the case of the nearest BS, it can be reduced to the well established result of the distance being *Rayleigh* distributed. We validate our results using a random node simulation and real Vodafone 3G network data, and go on to show how the distribution is tractable by deriving the average aggregate interference power.

## I. INTRODUCTION

### A. Background

Modelling the performance of cellular networks is vital to meeting performance targets. Femto-cells is universally seen as the solution to targeting poor coverage areas and providing increased spectrum re-use to boost capacity in popular areas [1]. The introduction of small-cells has made the computational modelling of large-scale cellular networks challenging due to the increased resolution required, and the need for indoor propagation models. Previously, a number of models that has attempted to analytically describe the performance of a heterogeneous cellular network. This includes models that account for a dominant interference source or approximate the contribution of far-field interference sources [2]. Stochastic geometry is a method that can yield a deterministic statistical performance of the whole network, as a function of key network and propagation environment properties [3]. Over the past few years, stochastic geometry has been utilised to analyse the performance of cellular and wireless mesh networks successfully [4].

### B. Stochastic Geometry

The paper considers a user $m$ at the origin who can detect the signal power of several neighbouring BSs. Given that the distance of each BS is large ($>$ several metres), the paper assumes a uniform pathloss distance exponent $\alpha$. As established in [3], the distribution of the distance from the origin to the nearest serving BS $R_1$ follows a Rayleigh $\left(\frac{1}{\sqrt{2\Lambda\pi}}\right)$ distribution, where $\Lambda$ denotes the BS density in the network. The relationship of the density to inter-site distance ($\delta$) is simply: $\Lambda \propto 1/\delta^2$.

What is lacking in literature is the extension of this distance distribution to the second, third and $n^{\text{th}}$ nearest BS. This is primarily because it is rare to consider a user attached to any BS other than the closest BS. However, recent research in advanced hand-over, cognitive radio, radio energy harvesting, and load balancing requires the knowledge of a general distance distribution. Given that the distance distributions in each tiers of a heterogeneous network are *independent*, we consider a generic single tier of BSs in this paper.

## II. A GENERAL DISTANCE DISTRIBUTION

**Theorem 1.** *The probability density function (pdf) of the distance between the $k^{\text{th}}$ nearest BS to the origin is:*

$$f_{R_k}(r;k) = \frac{2(\Lambda\pi)^k}{(k-1)!} r^{2k-1} e^{-\Lambda\pi r^2}, \quad k \geqslant 1. \qquad (1)$$

*Proof:* Recall that the probability of no BS is closer than r is given by $\mathbb{P}(0,r) = e^{-\Lambda\pi r^2}$ [3]. Hence, the probability of finding at least one BS within the distance of r is: $\mathbb{P}(n \geqslant 1, r) = 1 - e^{-\Lambda\pi r^2}$ and the probability of finding exactly one BS within the distance of r is $\mathbb{P}(1,r) = \Lambda\pi r^2 e^{-\Lambda\pi r^2}$. Therefore, the probability of finding at least two BSs within the distance of r is:

$$\begin{aligned} \mathbb{P}(n \geqslant 2, r) &= 1 - [\mathbb{P}(0,r) + \mathbb{P}(1,r)], \\ &= 1 - \left[e^{-\Lambda\pi r^2} + \Lambda\pi r^2 e^{-\Lambda\pi r^2}\right]. \end{aligned} \qquad (2)$$

Continuing in a similar manner, the probability of finding at least k BSs within the distance of r is given by:

$$\begin{aligned} \mathbb{P}(n \geqslant k, r) &= 1 - \sum_{m=0}^{k-1} \mathbb{P}(m, r), \\ &= 1 - e^{-\Lambda\pi r^2} - \Lambda\pi r^2 e^{-\Lambda\pi r^2} \cdots - \frac{(\Lambda\pi r^2)^{k-1}}{(k-1)!} e^{-\Lambda\pi r^2}. \end{aligned} \qquad (3)$$

The probability that the $k^{\text{th}}$ nearest BS to the origin is found in the annulus between the concentric circles radii r and $r + \Delta r$ is the difference between $\mathbb{P}(n \geqslant k, r + \Delta r)$ and $\mathbb{P}(n \geqslant k, r)$ and is expressed as

$$\mathbb{P}(k, r \sim r + \Delta r) = \mathbb{P}(n \geqslant k, r + \Delta r) - \mathbb{P}(n \geqslant k, r). \qquad (4)$$

The pdf of the distance from origin to the $k^{\text{th}}$ nearest BS is obtained by letting $\Delta r$ approach to the infinitesimal interval $\text{d}r$ and then differentiating Eq. (4) after substituting Eq. (3):

$$\begin{aligned} f_{R_k}(r;k) &= \frac{\text{d}\mathbb{P}(n \geqslant k, r)}{\text{d}r}, \\ &= \frac{\text{d}\left[1 - e^{-\Lambda\pi r^2} \cdots - \frac{(\Lambda\pi r^2)^{k-1}}{(k-1)!} e^{-\Lambda\pi r^2}\right]}{\text{d}r}, \\ &= \frac{2(\Lambda\pi)^k}{(k-1)!} r^{2k-1} e^{-\Lambda\pi r^2}, \quad k \geqslant 1. \end{aligned} \qquad (5)$$

∎

**Lemma 1.** *For* k = 1, *Eq.* (5) *reduces to* $2\Lambda\pi r e^{-\Lambda\pi r^2}$ *which is the same Rayleigh distribution of the distance to the nearest serving BS (*$R_1$*) derived in [3].*

**Lemma 2.** *Setting* $y = \Lambda\pi r^2$, $\Lambda\pi R_k^2$ *follows the classic form of gamma distribution with shape parameter* k *and scale parameter 1* $\left(\Lambda\pi R_k^2 \sim \Gamma(k,1)\right)$.

*Proof:*

$$f_{Y_k}(y;k) = \frac{d\mathbb{P}(n \geqslant k, r)}{dy} = \frac{d\mathbb{P}(n \geqslant k, r)}{2\Lambda\pi r dr},$$
$$= \frac{(\Lambda\pi)^{k-1}}{(k-1)!}r^{2k-2}e^{-\Lambda\pi r^2} = \frac{y^{k-1}}{(k-1)!}e^{-y}, \quad k \geqslant 1. \quad (6)$$

∎

**Lemma 3.** *Setting* $z = 2\Lambda\pi r^2$, $2\Lambda\pi R_k^2$ *can be proved to follow a chi-squared distribution with* 2k *degrees of freedom* $\left(2\Lambda\pi R_k^2 \sim \chi^2(2k)\right)$.

*Proof:*

$$f_{Z_k}(z;k) = \frac{d\mathbb{P}(n \geqslant k, r)}{dz} = \frac{d\mathbb{P}(n \geqslant k, r)}{4\Lambda\pi r dr},$$
$$= \frac{(2\Lambda\pi)^{k-1}}{2^k(k-1)!}r^{2k-2}e^{-\Lambda\pi r^2} = \frac{z^{k-1}}{2^k(k-1)!}e^{-\frac{z}{2}}, \quad k \geqslant 1. \quad (7)$$

∎

## III. VALIDATION AND APPLICATION

We now benchmark our distribution with two Monte-Carlo simulation models:

1) Random Model: Simulation with random BS placement;
2) London City 3G Network: Vodafone 3G Network with 92 macro-BSs spread over a 31 square kilometre area;

Fig. 1 shows the normalised histogram and theory ($\Lambda = 3$ per km$^2$) for distance from UE to BS, for both the general distribution (Fig. 1(b)) and the special case of closest BS (Fig. 1(a)). We can see that the derived distribution accurately described the closest BS (Rayleigh) and other BS (Gamma).

By knowing the distance distribution to any BS, we can derive useful identities such as the aggregate interference power. Let us define the aggregate interference ($N-1$ interfering BSs) as proportional to this random variable: $I = \sum_{j=2}^{N+1} P_j R_j^{-\alpha}$, where the constants of proportionality include transmit power ($P$) and pathloss constants ($\alpha$). The expected value of $I$ by definition is given by:

$$\begin{aligned}\mathbb{E}(I) &= \sum_{j=2}^{N+1} P_j \int_0^{+\infty} r^{-\alpha} \frac{2(\Lambda\pi)^j}{(j-1)!} r^{2j-1} e^{-\Lambda\pi r^2} \, dr, \\ &= \sum_{j=2}^{N+1} P_j \frac{(\Lambda\pi)^{\frac{\alpha}{2}} \Gamma(j-\frac{\alpha}{2})}{\Gamma(j)},\end{aligned} \quad (8)$$

where $\Gamma(x) = \int_0^{+\infty} e^{-t} t^{x-1} \, dt$.

## IV. SUMMARY AND FUTURE WORK

This paper has produced a novel distribution for the distance from an user to any base station. The general distribution is

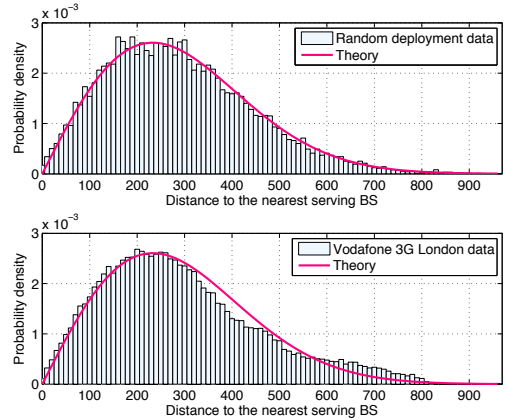

(a) Rayleigh distribution for distance to nearest (serving) BS

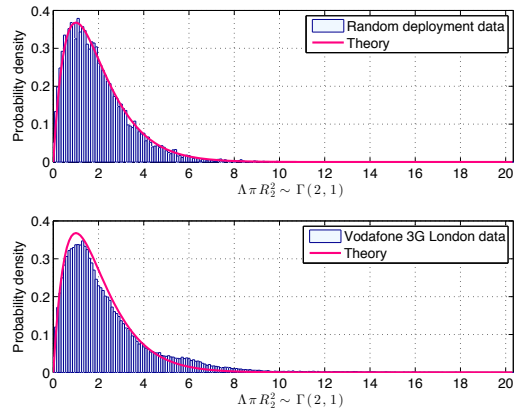

(b) Gamma distribution for random variable $Y_k$ (proportional to distance squared) to $k^{th}$ nearest BS (k = 2)

Fig. 1. Normalised histogram and theoretical pdf for distance from UE to BS for a) random model; and b) London city 3G network. Results are for the distance to the closest BS (Rayleigh distribution) and for the general random variable (Gamma distribution).

a more general form of the well known Rayleigh distribution for the nearest base station. The theory is proven using both random node placement and real cellular network data. The authors hope to extend this work to include the effect of antenna patterns and stochastic traffic loads into the analytical framework.